\documentstyle[aps,prd]{revtex}
\begin{document}
\title{Stochastic semiclassical gravity and fluctuations during inflation}

\author{Enric Verdaguer}
\address{Departament de F\'{\i}sica Fonamental,\\
Universitat de Barcelona, Av. Diagonal 647,\\
08028 Barcelona, Spain}
\maketitle

\begin{abstract}

Stochastic semiclassical gravity is a theory for the interaction
of gravity  with quantum matter fields which goes beyond the
semiclassical limit. The theory predicts stochastic fluctuations of the
classical gravitational field induced by the quantum fluctuations of the
stress energy tensor of the matter fields. Here
we use an axiomatic approach to introduce the Einstein-Langevin equations
as the consistent set of dynamical equations for a first order
perturbative correction to semiclassical gravity and review their main
features. We then describe the application of the theory in a simple
chaotic inflationary model, where the fluctuations of the
inflaton field induce stochastic fluctuations in the gravitational field.
The correlation functions for these gravitational fluctuations
lead to an almost Harrison-Zel'dovich scale invariant spectrum at large
scales, in agreement with the standard theories for structure
formation. A summary of recent results and other applications of the
theory is also given.

\end{abstract}


\section{Introduction}

Let us briefly summarize the road to stochastic semiclassical
gravity. It starts with quantum field theory in a curved spacetime which is
now a well understood and well defined theory at least for free fields
\cite{Birrell 84}. In this theory the gravitational field is the classical
field of general relativity, that is the metric of the spacetime, and the
quantum fields propagate in such a spacetime. Since the spacetime is now
dynamical it is not always possible to define a physically meaningful
vacuum state for the quantum field and when this is possible in some
``initial" times it is usually unstable, in the sense that it may differ
from the vacuum state at latter times, and spontaneous creation of
particles occurs. Applications of this in cosmology, such as particle
production in expanding Friedmann-Robertson-Walker models \cite{Parker},
and black hole physics, such as Hawking radiation \cite{Hawking}, are well
known. This is one aspect of the interaction of gravity with quantum matter
fields

Another aspect of this interaction is the back-reaction of the
quantum fields on the spacetime. Since the gravitational field couples to
the stress tensor of matter fields, the key object here is the
expectation value in a given quantum state of the stress
energy tensor of the quantum field, which is a classical observable.
However, since this object is quadratic in the field operator, which is
only well defined as a distribution on spacetime, it involves ill
defined quantities which translate into ultraviolet divergencies. To be
able to define a physically meaningful quantity a
regularization and a renormalization procedure is required. The ultraviolet
divergencies associated to the expectation value of the stress energy
tensor are also present in Minkowski spacetime, but in a curved
background the renormalization procedure is more sophisticated as it
needs to preserve general covariance. A regularization procedure which is
specially adapted to the curved background is the so called
point-splitting method \cite{Christensen}. The final expectation value of
the stress energy tensor using point splitting or any other reasonable
regularization technique is essentially unique, modulo some terms which
depend on the spacetime curvature and which are independent of the quantum
state. This uniqueness was essentially proved by Wald \cite{Wald} who
investigated the criteria that a physically meaningful expectation value
of the  stress energy tensor ought to satisfy.

Now the back-reaction problem may be formulated in terms of the so called
semiclassical Einstein equations. These are Einstein equations which have
the expectation value in some quantum state of the stress energy tensor as
a matter source. The back-reaction problem was investigated in
cosmology, in particular to see whether cosmological anisotropies
could be damped by back-reaction \cite{Lukash}. This was an earlier
attempt \cite{Misner} previous to inflation to explain why the universe is
so isotropic at present.

Another step forward in the back-reaction problem
was the use of effective action methods \cite{Hartle}
so familiar in quantum field
theory. These methods were of great help
in the study of cosmological anisotropies since they allowed the
introduction of familiar perturbative treatments into the subject. The
most common effective action method, however, led to equations of motion
which were not real because they were taylored to compute transition
elements of quantum operators rather than expectation values. Fortunately
the appropriate technique had already been developped by Schwinger
and Keldysh \cite{CTP1} in the so called Closed Time Path (CTP) or in-in
effective action method, and was soon adapted to the gravitational context
\cite{CTP2}. These techniques were then applied to different problems in
the cosmological context incuding the effects of arbitrary perturbations
on homogeneous backgrounds \cite{CTP3}. As a result one was now able to
deduce the semiclassical Einstein equations by the CTP functional
method: starting with an action for the interaction of gravity with
matter fields, treating the matter fields as quantum fields and the
gravitational field at tree level only.

The semiclassical Einstein equations have limitations, it is clear that
even outside Planck scales if fluctuations on the expectation value
of matter fields are large, and that depends on the quantum state, the
semiclassical equations should break down \cite{Waldfluc,Fordfluc,Kuo
and Ford}. One expects, in fact, that a better approximation would
describe the gravitational field in a probabilistic way. In other words,
that the semiclassical equations should be substituted by some
Langevin-type equations with a stochastic source that describes the quantum
fluctuations. A significant step in this direction was made by Hu \cite{Hu
89} who proposed to view the back-reaction problem in the framework of an
open quantum system, where the quantum fields are seen as the
``environment" and the gravitational field is seen as the ``system".
Following this proposal a systematic study of the connection between
semiclassical gravity and open quantum systems resulted in the
development of a framework were semiclassical Einstein-Langevin equations
could be derived \cite{SEL1}. The key technical factor to most of these
results was the use of the influence functional method of Feynman and
Vernon \cite{Feynman} for the description of the system-environment
interaction when only the state of the system is of interest. The CTP
method for open systems involves, in fact, the influence functional.

However although several Einstein-Langevin equations were derived, the
results were somewhat formal and some concern could be raised on the
physical reality of the solutions of the stochastic equations for the
gravitational field. This is related to the issue of the environment
induced quantum to classical transition. In fact, for the existence of a
semiclassical regime for the dynamics of the system one needs two
requeriments, in the language of the consistent histories formulation of
quantum mechanics \cite{Griffiths}. The first is decoherence, which
guarantees that probabilities can be consistently assigned to histories 
describing the evolution of the system, and the second is that these
probabilities should peak near histories which correspond to solutions of
classical equations of motion. The effect of the environment is crucial on
the one hand to provide decoherence \cite{Zurek} and on the other hand  to
produce both dissipation and noise to the system through back-reaction,
thus inducing a semiclassical stochastic dynamics on the system.
As shown by Gell-Mann and Hartle \cite{Gell-Mann} in an open quantum
system stochastic semiclassical equations are
obtained after a coarse graining of the environmental degrees of freedom
and a further coarse graining in the system variables. That this mechanism
could also work for decoherence and classicalization of the metric field
was not so clear lacking a quantum description of the gravitational field,
and the analogy could be made only formaly \cite{Martin 99a}.

Thus, an  axiomatic approach to the Einstein-Langevin equations which was
independent of the open system analogy was suggested:  it was based on the 
formulation of consistent dynamical equations for a perturbative correction
to semiclassical gravity able to account for the lowest order stress
energy fluctuations of matter fields \cite{Martin 99b}. It was later
shown that these same equations could be derived, in this general, case
from the influence functional of Feynman and Vernon in which, the
gravitational field is treated at tree level and and quantum fields are
quantized, the first being, in fact, the ``system" and the seconds the
``environment" \cite{Martin 99c}.

Here we review some of these developments in semiclassical stochastic
gravity and also some of its applications. In section \ref{2} a brief
sketch of semiclassical gravity is given. In section \ref{3} the
axiomatic approach to the Einstein-Langevin equations is discussed. To
illustrate the relation between the semiclassical, stochastic
semiclassical and quantum theories, a simplified model of linear gravity
is used. Throughout this section we use a simplified notation, avoiding
tensorial indices when possible and emphasizing the conceptual aspects. In
section \ref{4} an important application of stochastic gravity is
discussed in some detail, it concerns the computation of the two-point
correlations of the metric perturbations induced by the fluctuation in the
stress energy tensor of the inflaton field during inflation. The results
agree with the standard results but the present method in which the
gravitational field and the matter fields are treated separately may have
some advantages over the methods were both the metric and matter
perturbations are treated on the same footing from the start. Finally, in
section \ref{5} we summarize our result and briefly discuss other
applications. We should mention that an extensive and stimulating review
of stochastic semiclassical gravity is given in ref. \cite{Hu 99}, it
includes the suggestive idea of considering classical general relativity
it terms of collective variables and of viewing semiclassical gravity as
mesocopic physics.

\section{Semiclassical gravity}
\label{2}

Semiclassical gravity is a theory which describes the interaction of the
gravitational field assumed to be a classical field with matter fields
which are quantum. It is supposed to be some limit of the still unknown
theory describing the interaction of quatum gravity with quantum
fields. Due to the lack of the quantum theory, the semiclassical limit
cannot be rigurously derived. However, it can be formally derived in
several ways. One of them is the leading-order $1/N$ approximation of
quantum gravity \cite{Hartle and Horowitz}, where $N$ is the number of
independent free quantum fields which interact with gravity
only, and where one keeps finite the value of $NG$, where $G$ is Newton's
gravitational constant. In this limit, after path integration one arrives
at a theory in which formally the gravitational field can be treated as a
c-number (i.e. is quantized at tree level) and the quantum fields are
fully quantized. If we call $g$ the metric tensor and $\hat\phi$ the scalar
field (for simplicity we consider just one scalar field)
one arrives at the semiclassical Einstein equation as the
dynamical equation for the metric $g$:
\begin{equation}
G_g=8\pi G \langle \hat T^R\rangle_g,
\label{1.1}
\end{equation}
where $\hat T=T[\hat\phi^2]$ is the stress energy tensor in a simplifyied
notation, which is quadratic in the field operator $\hat \phi$. This
operator, being the product of distribution valued operators, is ill
defined and needs to be regularized and renormalized, the $R$ in $\hat
T^R$ means that the operator has been renormalized. The angle brackets on
the right hand side mean that the expectation value of the stress tensor
operator is computed in some quantum state, say $|\psi\rangle$, compatible
with the geometry described by the metric $g$. On the left hand side $G_g$
stands for the Einstein tensor for the metric $g$ together with the
cosmological constant term and other terms quadratic in the curvature
which are generally needed to renormalize the stress energy tensor
operator. The quantum field operator $\hat\phi$ propagates in the
background defined by the metric $g$, it thus satisfies a Klein-Gordon
equation, which we write also schematically as
\begin{equation}
\Box_g \hat\phi=0,
\label{1.2}
\end{equation}
where $\Box_g$ stands for the D'Alambert operator in the background of
$g$. Equation (\ref{1.1}) is the semiclassical Einstein equation, it is
the dynamical equation for the metric tensor $g$ and describes the
back-reaction of the quantum matter fields on the geometry. A solution of
semiclassical gravity consists of the set $(g,\hat \phi,|\psi\rangle)$
where $g$ is a solution of (\ref{1.1}), $\hat \phi$ is a solution
of (\ref{1.2}) and $|\psi\rangle$ is the quantum state in which the
expectation value of the stress energy tensor in (\ref{1.1}) is computed.

For a free quantum field this thery is robust in the sense that it is 
consistent and fairly well understood. Note that it is in some sense
unique as a theory where the gravitational field is classical. In
fact, the (classical) gravitational field interacts with matter fields
through the stress energy tensor, and the only reasonable c-number stress
energy tensor that one may construct \cite{Wald} with the operator
$\hat T$ is just the right hand side of (\ref{1.1}).
However the scope and limits of the theory are not so well understood as
a consequence of the lack of the full quantum theory. It is assumed that
the semiclassical theory should break down at Planck scales, which is when
simple order of magnitude estimes suggest that the quantum effects of
gravity cannot be ignored: the gravitational energy of a quantum
fluctuation of energy in a Planck size region, determined by the
Heisenberg uncertainty principle, are of the same order of magnitude.

There is also another situation
when the semiclassical theory should break down, namely, when the
fluctuations of the stress energy tensor are large. This has been
emphasized by Ford and collaborators. It is less clear how to quatify what
a large fluctuation here means and some criteria have been proposed
\cite{Kuo and Ford,Phillips and Hu}. Generally this depends on the
quantum state and may be illustrated by the example used in ref.
\cite{Fordfluc} as follows.

Let us assume a quantum state formed by an isolated system which consists
of a superposition with equal amplitude of one configuration with mass
$M_1$ and another with mass $M_2$. Semiclassical
theory as described in (\ref{1.1}) predicts that the gravitational field
of this system is produced by the average mass $(M_1+M_2)/2$, that is
a test particle will move on the background spacetime produced by such a
source. However one would expect that if we send a succesion of test
particles to probe the gravitational field of the above system half of the
time they would react to the field of a mass $M_1$ and the other half to
the field of a mass $M_2$. If the two masses differ substantially the two
predictions are clearly different, note that the fluctuations in mass of
the quantum state is of the order of $(M_1-M_2)^2$.
Although the previous example is suggestive a word of caution should be
said in order not to take it too literaly. In fact, if the previous masses
are macroscopic the quantum system decoheres very quickly \cite{Zurek} and
instead of a pure quantum state it is described by a density matrix which
diagonalizes in a certain pointer basis. Thus for observables associated
to this pointer basis the matrix density
description is  equivalent to that provided by a statistical ensemble. In
any case, however, from the point of view of the test particles the
predictions differ from that of the semiclassical theory.

\section{Einstein-Langevin equation}
\label{3}

The purpose of
semiclassical stochastic gravity is to be able to deal with the situation
of the previous example in which the predictions of
the semiclassical theory may be inacurate. Consequently, our first point is
to characterize the quantum fluctuations of the stress energy tensor.

The physical observable that measures these fluctuations is
$\langle \hat T^2\rangle-\langle \hat T\rangle^2$. To make this
more precise let us introduce the tensor operator $\hat t=\hat T- \langle
\hat T\rangle\hat I$, where $\hat I$ is the identity operator, then we
introduce the {\it{noise kernel}} as the four index bi-tensor defined as
the expectation value of the anticommutator of the operator $\hat t$:
\begin{equation}
N(x,y)={1\over2}\langle\{ \hat t(x),\hat t(y)\} \rangle_g.
\label{1.3}
\end{equation}
The subindex $g$ here means that this expectation value in taken in the
background metric which is a solution of the semiclassical equation
(\ref{1.1}). An important property of the symmetric bi-tensor $N(x,y)$ is
that it is finite  because the tensor operator $\hat t$ is finite since the
ultraviolet divergencies of $\hat T$ are cancelled by the substraction of
$\langle \hat T\rangle$. Since the operator $\hat T$ is selfadjoint
$N(x,y)$, which is the expectation value of an anticommutator, is real and
positive semi-definite. This last property allows for the introduction of a
classical Gaussian stochastic tensor $\xi$ defined by
\begin{equation}
\langle\xi(x)\rangle_c=0,\ \ \ \langle\xi(x)\xi(y)\rangle_c=N(x,y).
\label{1.4}
\end{equation}

This stochastic tensor is symmetric $\xi_{\mu\nu}=\xi_{\nu\mu}$ and
divergenceless, $\nabla^\mu\xi_{\mu\nu}=0$, as a consequence of the fact
that the stress tensor operator is divergenceless. The subindex $c$ means
that the expectation value is just a classical average. Note that we
assume that $\xi$ is Gaussian just for simplicity in
order to include the main effect. The idea now is simple we want to modify
the semiclassical Einstein equation (\ref{1.1}) by introducing a linear
correction to the metric tensor $g$, such as $g+h$, which accounts
consistently for the fluctuations of the stress energy tensor. The simplest
equation is,
\begin{equation}
G_{g+h}=8\pi G (\langle \hat T^R\rangle_{g+h}+\xi),
\label{1.5}
\end{equation}
where $g$ is assumed to be a solution of equation (\ref{1.1}). This
stochastic equation must be thought of as a linear equation for the metric
perturbation $h$ which will behave consequently as a stochastic field
tensor. Note that the tensor $\xi$ is not a dynamical source, since
it has been defined in the background metric $g$ which is solution of the
semiclassical equation. Note also that this source is divergenceless with
respect to the metric, and it is thus consistent to write it on the right
hand side of the Einstein equation. This equation is gauge
invariant with respect to diffeomorphisms defined by any field on the
background spacetime \cite{Martin 99b}. If we take the statistical average
equation (\ref{1.5}) becomes just the semiclassical
equation for the metric $g+h$ where now the expectation value of $\hat T$
is taken in the perturbed spacetime.

The stochastic equation (\ref{1.5}) is known as the Einstein-Langevin
equation. It is linear in $h$, thus its solutions can be written as the sum
of a solution of the homogeneous equation plus a stochastic
part: $h=h^h+h^s$. The equation predicts that the gravitational field
has stochastic fluctuations over the background $g$. The correlation
function for the gravitational field is simply given by $\langle
h^s(x)h^s(y)\rangle_c$. This is the physically most relevant
observable, to find it requires to solve the Einsten-Langevin
equation and to know the noise kernel $N(x,y)$. Note that the noise
kernel should be thought of as a distribution function, the limit of
coincidence points has meaning only in the sense of distributions.
Explicit expressions of this kernel in terms of the two point Wightman
functions is given in \cite{Martin 99b}, expression based on
point-splitting methods have also been given in \cite{Roura 00a,Phillips
and Hu 00}.

This stochastic theory goes beyond semiclassical gravity
in the following sense. The
semiclassical theory, which is based on the expectation value of the stress
energy tensor, carries information on the field two-point correlations
only, since $\langle \hat T\rangle$ is quadratic in the field operator
$\hat\phi$. The stochastic semiclassical theory on the other hand, is based
on the noise kernel (\ref{1.3})  which is quartic in the field. It is
thus clear that it carries information beyond the semiclassical theory.
In some sense the theory
represents a middle road between the semiclassical and the quantum theory.
It does not carry information, however, on the graviton-graviton
interaction.

\subsection{A simplified model}

To clarify this
point it is useful to introduce a toy model which 
would describe exactly a linear theory such as electromagnetism, and
captures some essential features of linearized gravity. Let
us assume that the gravitational equations are described by the linear
equations for the field $h$, with a source $T[\phi^2]$. The semiclassical
equations now correspond to
\begin{equation}
\Box h=\langle \hat T\rangle,
\label{1.6}
\end{equation}
where now $\hat T$ depends on the field operator $\hat \phi$. Note that
this theory does not correspond to linearized semiclassical gravity
around the Minkowski background since in that case $\langle
\hat T\rangle=0$ assuming that $T$ does not depend on $h$
(in the linearized semiclassical theory $\langle
\hat T\rangle$ is, in fact, linear $h$ \cite{Horowitz}). The model,
however, can be extended to linearized gravity \cite{Roura 00b}.

The solutions of this equation may be
written in terms of the retarded propagator $G_{xy}$ as
\begin{equation}
h_x=h_x^0+\int G_{xx'}\langle \hat T_{x'}\rangle,
\label{1.7}
\end{equation}
where $h_x^0$ is the homogeneous solution which is determined by the
initial conditions.

Let us now consider the quantum theory which, in
Heisenberg representation, may be written as
\begin{equation}
\Box \hat h=\hat T.
\label{1.8}
\end{equation}
The solutions of this equation may be written as 
\begin{equation}
\hat h_x=\hat h_x^0+\int G_{xx'}\hat T_{x'},
\label{1.9}
\end{equation}
and one may compute the two point quatum correlation function as
\begin{equation}
\langle \hat h_x\hat h_y\rangle = \langle \hat h^0_x\hat h^0_y\rangle +
\int\int G_{xx'}G_{yy'} \langle\hat T_{x'}\hat T_{y'}\rangle,
\label{1.10}
\end{equation}
where the expectation value is taken in the quantum state in which both
fields $\phi$ and $h$ have been quantized, and we have used that
for the free field $\langle \hat h^0\rangle=0$.

At this point we may now compare with the stochastic theory as described
by equation (\ref{1.5}) which, in this simplified model, may be written as  
\begin{equation}
\Box  h=\langle \hat T\rangle +\xi,
\label{1.11}
\end{equation}
where $\xi$ is a Gaussian stochastic source defined by
$\langle \xi_x\rangle_c=0$ and $\langle \xi_x\xi_y\rangle_c=
\langle\hat T_x\hat T_y\rangle-\langle\hat T_x\rangle
\langle\hat T_y\rangle$, where the expectation values on the right hand
side are defined in a given state of the field $\hat \phi$ and the
subscript $c$ on the left hand sides means statistical average, see
(\ref{1.3}) and (\ref{1.4}). Now the solution of this equation may be
written in terms of the retarded propagator as,
\begin{equation}
h_x=h_x^0+\int G_{xx'}\left( \langle \hat T_{x'}\rangle +\xi_{x'}\right) ,
\label{1.12}
\end{equation}
from where the two point correlation function for the classical field $h$,
after using the definition of $\xi$ and that $\langle h^0\rangle_c=0$, is
given by
\begin{equation}
\langle  h_x h_y\rangle_c = \langle  h^0_x h^0_y\rangle_c +
\int\int G_{xx'}G_{yy'} \langle\hat T_{x'}\hat T_{y'}\rangle.
\label{1.13}
\end{equation}

Comparing (\ref{1.10}) with (\ref{1.13}) we see that the respective second
terms on the right hand side are identical provided the expectation
values are computed in the same quantum state for the field $\hat \phi$,
note that we have assumed that $T$ does not depend on $h$. The fact that
the field $h$ is also quantized in (\ref{1.10}) does not change the
previous statement. In the real theory of gravity $T$, in fact, depends
also on $h$ and then the previous statement is only true approximately,
i.e  perturbatively in $h$. The nature of the first terms on the
right hand sides of equations (\ref{1.10}) and (\ref{1.13}) is different:
in the first case it is the two point expectation value of the free
quatum field $\hat h_0$ whereas in the second case it is the average of the
two point classical average of the homogeneous field $h_0$, which depends
on the initial contions. Now we can still make these terms to be equal to
each other if we assume for the homogeneous field $h$ a distribution of
initial conditions such that $\langle  h^0_x h^0_y\rangle_c =   \langle
\hat h^0_x\hat h^0_y\rangle$. Thus, under this assumption on initial
conditions for the field $h$ the two point correlation function of
(\ref{1.13}) equal the quantum expectation value of (\ref{1.10}) exactly.
Thus in a linear theory as in the model just described one may just use
the statistical description given by (\ref{1.11}) to compute the
quantum two point function of equation (\ref{1.10}). Of course, the
statistical description is not able to account for graviton-graviton
effects which go beyond the linear approximation in $\hat h$.

\subsection{Functional approach}

To end this section we should mention that the Einstein-Langevin equation
(\ref{1.5}) may also be formaly derived using the CTP
functional method \cite{Martin 99a}. As remarked in the introduction
the CTP functional was introduced by Schwinger \cite{CTP1} to
compute expectation values. One just considers the
interaction of the gravitational field $g$ at tree level and of the quantum
field $\phi$ fully quantum. Then the effective action for the
gravitational field is derived after integrating out the degrees of
freedom of the quantum field, and the CTP influence action reduces
basically to the Feynman and Vernon influence functional \cite{Feynman}
used in quantum open systems. Here the system is the gravitational field
and the environment is the quantum field. The stochastic terms for the
gravitational field are found by suitably interpreting some pure
imaginary term which appear in the influence action. These terms are
closely connected to Gell-Mann and Hartle decoherence functional
\cite{Gell-Mann} used to study decoherence and classicalization in open
quantum systems. The net result of these
analogies is that the interaction with the environmennt induces
fluctuations in the system dynamics. It is precisely the
noise kernel introduced in (\ref{1.3}) that accounts for this effect.

\section{Gravitational fluctuations during inflation}
\label{4}

An important application of stochastic semiclassical
gravity is the derivation of the
cosmological perturbations generated during inflation \cite{Roura 00a}.
Let us consider the Lagrangian density for the inflaton field $\phi$ of
mass $m$
\begin{equation}
{\cal L}(\phi)={1\over 2}(\partial\phi)^2 +
{1\over 2}m^2\phi^2,
\label{1.14}
\end{equation}
which is the basis of the simplest chaotic inflationary model \cite{Linde
90}. The conditions for the existence of an inflationary period, which is
characterized by an accelerated expansion of the spacetime, is that the
value of the field averaged over a region with the typical size of the
Hubble radius is higher than the Planck mass $m_P$. This is because in
order to solve the cosmological horizon and flatness problem more than 60
e-folds of expansion are needed, to achieve this the scalar field should
begin with a value higher than $3m_P$. Furthermore, as we will see, the
large scale anisotropies measured \cite{Smoot 92} restrict the inflaton
mass to be of the order of $10^{-6}m_P$.

We want to study the metric perturbations produced by the stress tensor
fluctuations of the inflaton field on the homogeneous background of a flat
Friedmann-Robertson-Walker model, described by the cosmological scale
factor $a(\eta)$, where $\eta$ is the conformal time, which is driven by
the homogenous inflaton field $\phi(\eta)=\langle\hat\phi\rangle$. Thus we
write the inflaton field in the following form
\begin{equation}
\hat\phi=\phi(\eta)+ \hat\varphi (x),
\label{1.15}
\end{equation}
where $\hat\varphi (x)$ corresponds to a free massive quantum scalar field
with zero expectation value on the homogeneous background metric:
$\langle\hat\varphi\rangle_g=0$. 
Restrincting ourselves to scalar-type perturbations the perturbed metric
$\tilde g=g+h$ can be written in the longitudinal gauge as,
\begin{equation}
ds^2=a^2(\eta)[-(1+2\Phi(x))d\eta^2+(1-2\Psi(x))\delta_{ij}dx^idx^j],
\label{1.16}
\end{equation}
where the metric perturbations $\Phi(x)$ and $\Psi(x)$ correspond to
Bardeen's gauge invariant variables \cite{Bardeen}.

The Einstein-Langevin
equation as described in the previous section is gauge invariant, and thus
we can work in a desired gauge and then extract the gauge invariant
quantities. The Einstein-Langevin
equation (\ref{1.5}) reads now:
\begin{equation}
G^{(0)}-8\pi G\langle\hat T^{(0)}\rangle_g+
G^{(1)}(h)-8\pi G\langle\hat T^{(1)}(h)\rangle_g= 8\pi G\xi,
\label{1.17}
\end{equation}
where the two first terms cancel, that is
$G^{(0)}-8\pi G\langle\hat T^{(0)}\rangle_g=0$, as the background metric
satisfies the semiclassical Einstein equations. Here the subscripts $(0)$
and $(1)$ refer to functions in the background metric $g$ and linear
in the metric perturbation $h$, respectively. The stress
tensor operator $\hat T$ for
the minimally coupled inflaton field in the perturbed metric $\tilde
g=g+h$ is:
\begin{equation}
\hat
T_{\mu\nu}=\partial_{\mu}\hat\phi\partial_{\nu}\hat\phi+{1\over2}\tilde
g_{\mu\nu} (\partial_{\rho}\hat\phi\partial^{\rho}\hat\phi+
m^2\hat\phi^2).
\label{1.18}
\end{equation}

Now using the decomposition of the scalar field into its
homogeneous and inhomogeneous part, see  (\ref{1.15}), and the metric
$\tilde g$ into its homogenous background $g$ and its perturbation $h$,
the renormalized expectation value for the stress tensor can be written
as 
\begin{equation}
\langle \hat T^R[\tilde g]\rangle=
\langle \hat T[\tilde g]\rangle_{\phi\phi}+
\langle \hat T[\tilde g]\rangle_{\phi\varphi}+
\langle \hat T^R[\tilde g]\rangle_{\varphi\varphi},
\label{1.19}
\end{equation}
where only the homogeneous solution for the scalar field contributes to
the first term. The second term is proportional to
$\langle\hat\varphi[\tilde g]\rangle$ which is not zero because the field
dynamics is considered on the perturbed spacetime, i.e. this term includes
the coupling of the field with $h$. The last term corresponds to the
expectation value to the stress tensor for a free scalar field on the
spacetime of the perturbed metric.

We can now compute the noise kernel
$N(x,y)$ defined in equation (\ref{1.3}), which after using the previous
decomposition can be written as
\begin{equation}
\langle \{\hat t,\hat t\}\rangle[g]=
\langle \{\hat t,\hat t\}\rangle_{\phi\varphi}[g]+
\langle \{\hat t,\hat t\}\rangle_{\varphi\varphi}[g],
\label{1.20}
\end{equation}
where we have used the fact that $\langle\hat\varphi\rangle_g=0
=\langle\hat\varphi\hat\varphi\hat\varphi\rangle_g$ for Gaussian states on
the background geometry. We have considered the vacuum state to be the
Euclidean vacuum which is preferred in the de Sitter
background, and this state is Gaussian. In the above equation the first
term is quadratic in $\hat\varphi$ whereas  the second one is quartic,
both contributions to the noise kernel are separately conserved since
both $\phi(\eta)$ and $\hat\varphi$ satisfy the Klein-Gordon field
equations on the background spacetime. Consequently, the two terms can be
considered separately. On the other hand if one treats $\hat \varphi$
as a small perturbation the second term in (\ref{1.20}) is of lower order
than the first and may be neglected consistently, this corresponds to
neglecting the last term of (\ref{1.19}). The stress tensor fluctuations
due to a term of that kind were considered in ref. \cite{Roura 99}.

We can now write down the Einstein-Langevin equations (\ref{1.17}). It is
easy to check that the {\it space-space} components coming from the stress
tensor expectation value terms and the stochastic tensor are diagonal,
i.e. $\langle\hat T_{ij}\rangle=0= \xi_{ij}$ for $i\not= j$. This, in
turn, implies that the two functions characterizing the scalar metric
perturbations are equal: $\Phi=\Psi$ in agreement with ref.
\cite{Mukhanov 92}. The equation for $\Phi$ can be obtained from the
$0i$-component of the Einstein-Langevin equation, which in Fourier
space reads
\begin{equation}
2ik_i(H\Phi_k+\Phi'_k)= {8\pi\over m_P^2}\xi_{k\, 0i},
\label{1.21}
\end{equation}
where $k_i$ is the comoving momentum component associated to the comoving
coordinate $x^i$. Here primes denote derivatives with respect to the
conformal time $\eta$ and $H=a'/a$. A non-local term of dissipative
character which comes from the second term in (\ref{1.19}) should also
appear on the left hand side of equation (\ref{1.21}), but since we are
mainly interested in the fluctuating part we have ignored this term.
To solve this equation, whose left hand side comes from the linearized
Einstein tensor for the perturbed metric \cite{Mukhanov 92}, we need the
retarded propagator for the gravitational potential $\Phi_k$,
\begin{equation}
G_k(\eta,\eta')= -i {4\pi\over k_i m_P^2}\left( \theta(\eta-\eta')
{a(\eta')\over a(\eta)}+f(\eta,\eta')\right),
\label{1.22}
\end{equation}
where $f$ is a homogeneous solution of (\ref{1.21}) related to the initial
conditions chosen. For intance, if we take
$f(\eta,\eta')=-\theta(\eta_0-\eta')a(\eta')/a(\eta)$ the solution would
correspont to ``turning on" the stochastic source at $\eta_0$.

The
correlation function for the metric perturbations is now given by
\begin{equation}
\langle\Phi_k(\eta)\Phi_{k'}(\eta')\rangle_c= (2\pi)^2\delta(\vec
k+\vec k')\int^\eta d\eta_1\int^{\eta'}d\eta_2 G_k(\eta,\eta_1)
G_{k'}(\eta',\eta_2)
\langle\xi_{k\ 0i}(\eta_1)\xi_{k'\ 0i}(\eta_2)\rangle_c .
\label{1.23}
\end{equation}
The correlation function for the stochastic source , which is connected to
the stress tensor fluctuations through the noise kernel is given by,
\begin{equation}
\langle\xi_{k\ 0i}(\eta_1)\xi_{-k\ 0i}(\eta_2)\rangle_c= {1\over2}
\langle\{\hat t^k_{0i}(\eta_1,\hat
t^{-k}_{0i}(\eta_2)\}\rangle_{\phi\varphi}= {1\over2}
k_ik_i\phi'(\eta_1)\phi'(\eta_2)G_k^{(1)}(\eta_1,\eta_2),
\label{1.24}
\end{equation}
where $G_k^{(1)}(\eta_1,\eta_2)=\langle\{\hat\varphi_k(\eta_1),
\hat\varphi_{-k}(\eta_2)\}\rangle$ is the $k$-mode Hadamard function for a
free minimally coupled scalar field which is in the
Euclidean vacuum on the de Sitter background.

It is useful to compute the Hadamard function for a massless field and
consider a perturbative expansion in terms of the dimensionless parameter
$m/m_P$. Thus we consider $\bar
G_k^{(1)}(\eta_1,\eta_2)= a(\eta_1)a(\eta_2)G_k^{(1)}(\eta_1,\eta_2)=
\langle 0|\{\hat y_k(\eta_1),\hat y_{-k}(\eta_2)\}|0\rangle=
2{\cal R}\left(u_k(\eta_1)u_k^*(\eta_2)\right)$ 
with $\hat y_k(\eta)= a(\eta)\hat\varphi_k(\eta)=
\hat a_k u_k(\eta)+\hat a_{-k}^\dagger u_{-k}^*(\eta)$ and where
$u_k=(2k)^{-1/2}e^{ik\eta}(1-i/\eta)$ are the positive
frequency $k$-mode for a massless minimally coupled scalar field
on a de Sitter background, which define the
Euclidean vacuum state: $\hat a_k|0\rangle=0$ \cite{Birrell 84}. 

The background geometry, however, is not exactly that of de Sitter
spacetime, for which $a(\eta)=-(H\eta)^{-1}$ with $-\infty <\eta< 0$.
One can expand in terms of the ``slow-roll" parameters and assume that to
first order $\dot\phi(t)\simeq m_P^2(m/m_P)$, where $t$ is the physical
time. The correlation function for the metric perturbation
(\ref{1.23}) is the computed, see ref. \cite{Roura 00a} for details. The
final result, however, is very weakly dependent on the initial conditions
as one may understand from the fact that the accelerated expansion of de
quasi-de Sitter spacetime during inflation erases the information about
the initial conditions. Thus one may take the initial time to be
$\eta_0=-\infty$ and obtain to lowest order in $m/m_P$ the expression
\begin{equation}
\langle\Phi_k(\eta)\Phi_{k'}(\eta')\rangle_c\simeq
8\pi^2\left( {m\over m_P}\right)^2 k^{-3}(2\pi)^3\delta(\vec k+\vec k')
\cos k(\eta-\eta').
\label{1.25}
\end{equation}

From this result two main conclusions are derived. First, the prediction
of an almost Harrison-Zel'dovich scale-invariant spectrum for large scales,
i.e. small values of $k$. Second, since the correlation function is of
order of $(m/m_P)^2$ a severe bound to the mass $m$ is imposed by the
gravitational fluctuations derived from the small values of the Cosmic
Microwave Background (CMB) anisotropies detected by COBE. This
bound is of the order of $(m/m_P)\sim 10^{-6}$ \cite{Smoot 92,Mukhanov 92}.
One possible advantge of the Einstein-Langevin approach to the
gravitational fluctuations in inflaton over the approach based on the
quantization of the linear perturbations of both the metric and the
inflaton field \cite{Mukhanov 92}, is that an exact treatment of the
inflaton quantum fluctuations is in principle possible, keeping the metric
perturbations to linear order. On the other hand although the
gravitational fluctuations are here assumed to be classical, the
correlation functions obtained correspond to the quantum
expectation values of the quantum metric perturbations \cite{Calzetta et
al 00,Roura 00b}, at least in the linear regime. This means that even in
the absence of decoherence the fluctuations predicted by the
Einstein-Langevin equation, whose solutions do not describe the
actual dynamics of the gravitational field any longer, still give the
correct quantum two-point functions.

\section{Summary and outlook}
\label{5}

We have reviewed the semiclassical theory of gravity as the theory of
the interaction of classical gravity with quantum matter fields. The most
important equations in this theory are the semiclassical Einstein
equations (\ref{1.1}) which describe the back-reaction of the
gravitational fluctuations in its interaction with the quantum fields. We
noticed that the theory may seriously fail when the fluctuations on the
stress energy tensor of the quantum fields are
important. We have then sought an axiomatic approach by which the
semiclassical equations can be corrected in order to take into account
those fluctuations. These equations turn out to be uniquely defined and
are the Einstein-Langevin equations (\ref{1.5}) which are linear in the
metric perturbations $h$ over the semiclassical background. These
equations predict stochastic fluctuations in the metric  perturbations
induced by the stress tensor fluctuations described by the noise kernel
(\ref{1.3}).

We have also noticed that the Einstein-Langevin equations can be
formally derived from the open quantum system paradigm, in which the
gravitational fluctuations $h$ and the quantum field interact when the
interest is in the dynamics of the gravitational field. Thus treating the
quantum field as the ``environment" and the gravitational field as the
``system". The mathematical tools to carry out this approach are the CTP
functional method and, in this context, its closely related Feynman
and Vernon influence functional.

We have finally used the stochastic theory in the inflationary
cosmological context. We have computed the two-point correlation
functions of the metric fluctuations during a quasi-de
Sitter expansion induced by the stress tensor fluctuations of the
inflaton field. The results are in agreement with other approaches to the
same problem \cite{Mukhanov 92}, an approximate Harrison-Zel'dovich
spectrum is predicted. We noticed that in our approach the quantum fields
and the gravitational fields are treated separately, and this may
have some advantages to go one step further and consider the quantum
field fully, not just to linear order.

In is worth mentioning that other applications of the stochastic theory
have been carried out and others are in progress. Thus, the fluctuations on
the ground state of semiclassical gravity, which consists of the
Minkowski metric and the quantum state in its vacuum state, have been
considered \cite{Martin 00}. The computation of the two-point correlations
of the linearized Einstein tensor indicate that a typical correlation
length is present.

Other applications of stochastic semiclassical gravity
to semiclassical cosmology have been performed \cite{Calzetta 97a}, some
including thermal fields \cite{Campos and Hu,Martin 99c}. It has been
shown that noise produced by a quantum  field on the cosmological scale
factor of an isotropic closed Friedmann-Robertson-Walker, in the presence
of a cosmological constant, may take the scale factor from a region where 
it is nearly zero to a region where it describes a de Sitter inflationary
era \cite{Calzetta 99}. Thus jumping over the barrier by activation, this
is the semiclassical analogue of the tunneling from nothing in quantum
cosmology \cite{Vilenkin} and gives yet another mechanism to produce
inflation.

An important application of stochastic gravity which is now beginning
is in the physics of black holes \cite{black holes,Hu 99}. In particular in
black hole thermodynamics, the stress tensor fluctuations near the black
hole horizon may induce fluctuations in the horizon area. The relevance of
this back-reaction effect in Hawking radiation has not been yet explored,
although preliminary investigations seem to indicate that Hawking
result should not be substantially different \cite{Ford2}. The
contribution of the horizon fluctuations to the black hole entropy
\cite{Sorkin} is another tantalizing issue that may deserve some attention
in the present context.

\acknowledgements

I am very grateful to Professors Gr. Tsagas and D. Papadopoulos for
giving me the oportunity to participate at the conference on {\sl
Applied and differential geometry, Lie algebras and general relativity},
and for their kind and genereous hospitality. I am also grateful to Rosario
Mart\'{\i}n and Albert Roura for their unvaluable contribution to the work
described here. I also thank Albert Roura for a critical reading of the
manuscript and many fruitful discussions. This work has been partially
supported by the CICYT Research Project No. AEN98-0431.

\end{document}